# High-Speed Signature Matching in Network Interface Device using Bloom Filters

Arun Kumar S P
Member, IEEE
arunkumarsp@ymail.com

*Abstract*— Network intrusion detection systems play a critical role in protecting the information infrastructure of an organization. Due to the sophistication and complexity of techniques used for the analysis they are commonly based on general-purpose workstations. Although cost-efficient, these general-purpose systems are found to be inadequate as they are unable to perform efficiently at high packet rates. The resulting packet loss degrades the system's overall effectiveness, as the analyzing capability of the system is reduced. It has been found that the performance of these sensors can be improved significantly by filtering out unwanted packets. This paper presents the design of a Programmable Ethernet Interface Card that is used to offload signature matching from software and thereby improve the detection ratio and performance of the system.

## I. INTRODUCTION

A Network Intrusion Detection System (NIDS) is a tool that collects network activity data and analyzes the information to determine whether there is an attack occuring. It tries to detect malicious activity such as denial of service attacks, port scans or even attempts to crack into computers by passively monitoring network traffic [1].

A variety of techniques exist to infer malicious network activity from the network traffic. The popular methods employ signature matching and anomaly detection. In signature matching, [2, 3] patterns that characterize well-known attacks are looked for in the packet payload. In anomaly detection [4, 5, 6, 7, 8], deviations from normal traffic characteristics are used to signal an intrusion. Regardless of the particular analysis technique employed, it is vital that the NIDS is able to process packets at the wire-speed. Any packet loss results in a proportional loss in intrusion detection effectiveness [9]. Being a passive device, a NIDS does not have the ability to reject connections or use the flow control mechanisms to adjust to the load. Attackers can use this weakness and deliberately avoid detection by overloading the NIDS sensor with malignant traffic. This causes high packet loss and an increase in the probability of a successful intrusion going undetected [10]. Therefore, the system must be designed to handle worst-case traffic scenarios.

Due to the sophistication and complexity of analysis performed by NIDS, they are commonly executed on general-purpose workstations. Unfortunately, it has been observed that modern general-purpose systems are inadequate as NIDS platforms because of their limitations in their capability to analyze high-bandwidth network traffic [11]. High interrupt rates and the interrupt handling architecture of the operating system make general purpose machines less effective for these kinds of applications. It has been found that the performance of an NIDS can be improved significantly by identifying and filtering away packets that are not of interest at the interface [12]. This can improve the performance of the system significantly.

To this end, an efficient signature matching algorithm was implemented on a FPGA-based Programmable Ethernet Interface Card (PEIC) that supports 10/100 Mbps Ethernet traffic. The PEIC was used as the platform to offload the signature matching done in software to the PEIC. Thus all irrelevant information is filtered away, thereby providing the host microprocessor sufficient computing resources to inspect the subset of the traffic that requires security analysis. The signature matching is performed at the PEIC using a concise, probabilistic data structure called Bloom Filter.

The paper begins with a survey of related work in Section II. Section III discusses the relevant theory behind the Bloom Filters. Section IV describes the overall hardware architecture and the implementation details. Section V evaluates the architecture and presents the results. Section VI concludes the paper.

## II. RETATED WORK

Previous works have evaluated network intrusion detection systems in an attempt to identify the various inadequacies in their performance. In [9] the authors have presented a simple methodology to measure and characterize the performance of general-purpose systems when used as network intrusion detection systems. A comparative study of six different general purpose machines was carried out and the results show that these systems are inadequate for even moderate network speeds.

The idea of using hardware-based systems to improve NIDS performance has also been investigated over the past several years. Paul Willmann, et al discusses the hardware and software mechanisms in the design of 10 Gigabit programmable network interface card in [13]. The concept of a Gigabit reconfigurable network interface card is discussed in [14]. Their architecture is based on Xilinx FPGAs and PowerPCs as embedded processors. In [15] the authors describe a framework for Network Intrusion Prevention system at gigabit rates, based on Intel's IXP2400 network processor. Authors in [16] present and evaluate a NIC-based network intrusion detection system. They considered embedding both signature detection algorithms as well anomaly detection algorithms in the interface card itself. Although this completely offloads the CPU, the algorithms used were simplified versions thus affecting the quality and detection rate. Authors in [17] have implemented NIC-based filtering for similar application, but this was limited to header information.

## III. BACKGROUND

The algorithm used for signature matching is based on Bloom Filters. It is a widely used data structure for content analysis. A Bloom Filter is an ingenious space-efficient randomized data-structure for concisely representing a set in order to support approximate membership queries. The space efficiency is achieved at the cost of a small, configurable probability of false positives. Bloom Filters were invented in the 1970's [18] and have been heavily used in database [19, 20] and networking applications [21, 22, 23]. This data structure was chosen for implementation due to its space efficiency and ease of hardware implementation.

A Bloom filter represents an n-element set $S=\{X_1, X_2 ... X_n\}$ by using a bit vector $B = B_1B_2....B_m$ of length m. Initially all the bits are set to 0. The filter uses k independent hash functions $h_1 ... h_k$ with range $\{1...m\}$, i.e. $h_i: X \rightarrow \{1... m\}$, $1 \leq i \leq k$. For optimal performance, each of the k hash functions should be a member of the class of universal hash functions [24]. This implies that these hash functions map each item in the universe to a random number that is uniform over the range $\{1 ... m\}$.

To start with, the Bloom Filter is 'programmed' to include all the set elements (Refer Figure 1). To store an element $x \in S$, the bits $h_i(x)$ are set for $1 \leq i \leq k$. To check if an element y is in S, one simply checks whether all $h_i(y)$ are set. If not, then clearly y is not a member of S. However, if all $h_i(y)$ are asserted, one cannot infer that element y is definitely in S. It is possible that by coincidence $h_1(y),...,h_k(y)$ are all set to 1. This situation is called *false–positive* and the probability that this occurs is called *false-positive rate*. Hence a Bloom filter does not yield a *false negative* but may yield a false positive, where it suggests that an element y is in S even though it is not.

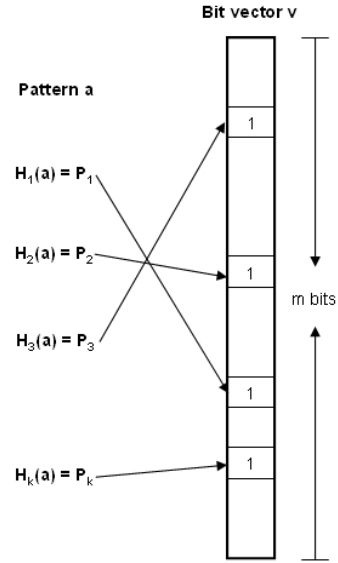

Figure 1. Bloom Filter programming. Different hash functions are calculated over the input. The hash values obtained are used as pointers to an m bit vector and are set to indicate membership.

If $V[]$ denotes the m bit vector, and $H$ denotes the hash function over the invariant fields $f_1,...f_n$, and k hash functions are used, then the Bloom Filter programming can be represented as

$$\forall i, \quad V[H_i(f_1...f_n)] \rightarrow 1 \qquad (2)$$

The pseudo code for programming the Bloom Filter is given below.

```
bloom_add(BLOOM *bloom, element s)
{
  for (k=0; k<bloom->nfuncs; n++) {
    SETBIT(bloom->a,bloom->hash[k](s));
  }
}
```

In the above code, *hash[]* contain the various k hash functions. This is called for every element s that needs to be stored.

Checking for an element can also be done in a similar way by verifying that all the locations in the bit vector pointed by the hash functions are set. Thus if Π denotes the bitwise AND operation and denotes the hash function over the invariant fields, the membership is asserted only if

$$\prod_{1}^{k} H_i(f_1,...f_n) = 1 \qquad (3)$$

The pseudo code for membership query to the Bloom Filter is given below.

```
bloom_check(BLOOM *bloom, element s)
{
    for (k=0; k<bloom->nfuncs; n++) {
    if (!(GETBIT(bloom->a, bloom->hash[k](s))))
            return NOT_MEMBER;
    }
    return MEMBER;
}
```

## IV. PROPOSED ARCHITECTURE

This section describes the architecture of the system. The objective is to prune all superfluous traffic at the interface, thus reduce the interrupt rate and the analysis overhead.

In general, NIDS software acquires packets directly from the network interface buffer using one of the various packet capture libraries like Pcap and BPF. Some systems allow usage of Linux netfilter kernel hooks for obtaining packets. The former approach is popular owing to its availability in various platforms. Figure 2 shows the normal data flow in a NIDS system.

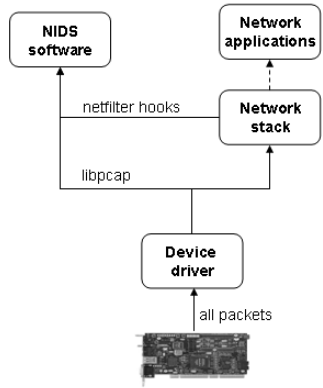

Figure 2.     Standard Flow

An interrupt is generated whenever the network card receives a packet. The interrupt service routine, which is part of the device driver fetches the packet and passes it on to the analysis software. The analysis software executes signature matching algorithms on the packet payload.

Figure 3 shows the modified flow using PEIC. The received packets are provisionally stored in the card buffers. Signature matching algorithm is applied to the packets and only packets that warrant further analysis are passed to the software. Due to the false positives caused due to the data structure a small number of packets are also sent to the host. But this rate is configurable and the architecture considerably reduces the interrupts that are generated by the interface. Moreover, as the number of packets that are sent to the host is reduced, the load on the NIDS software is lowered significantly.

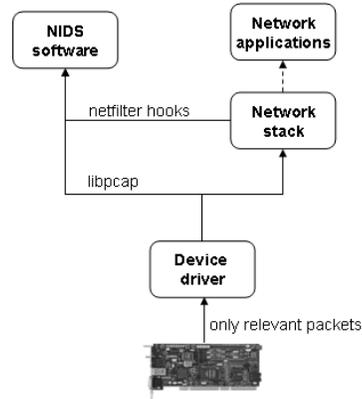

Figure 3.     PEIC Flow

A FPGA-based Programmable Ethernet Interface Card (PEIC) was used to handle 10/100 Ethernet traffic. The PEIC connects the PCI-compliant server to a Fast Ethernet network and is implemented for use in systems that support 32 bit wide PCI Local Bus operating at 33 MHz.

The block diagram of the PEIC is shown in Figure 4.

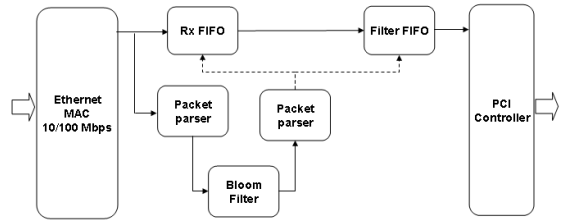

Figure 4.   Block Diagram

Altera's PCI Development Kit was used for development. It is based on a Cyclone II EP2C35F672 FPGA which houses all of the processing logic. External network interface is provided through a 10/100 Ethernet port with a SMSC MAC controller. Host interface is the standard 32 bit PCI bus connector. A RS-232 serial port is also provided for external configuration for the card. Figure 5 shows the card.

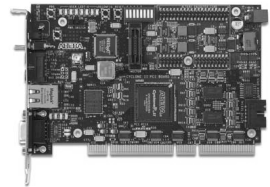

Figure 5.   Prototype card

The PEIC provides high flexibility, in addition to providing the basic tasks of a network interface card. The MAC interface is responsible for accepting packets from the line. The received packets go through a packet parser logic which extracts the payload information for content analysis. The data from the MAC is temporarily stored in a synchronizing FIFO until the filtering decision is made. The signature matching logic uses the payload information provided by the parser, compares it with an internally stored rule set and decides whether to allow the packet or not. Based on the match decision, the FIFO controller selectively transfers the data to the Filter FIFO. The host interface is provided through a standard PCI connector. A 32-bit 33 MHz Master/Target PCI controller is used for this purpose. Packet arrival indication is communicated through an interrupt. A PCI interrupt is generated whenever a packet is written into the Filter FIFO by setting a mailbox register in the PCI controller.

A device driver was also developed for Linux 2.4 kernel based systems. The driver supports programming of the programming of the Bloom Filter vector from the operating system.

## V. EVALUATION AND RESULTS

The following sections describe the experiments for evaluating the effects of filtering in software and hardware.

### A. Experimental Setup

For the experiments, a 3.2 GHz Pentium IV processor PC with 16 KB L1 cache, 1024 KB L2 cache and 512 MB of main memory was used for running the NIDS software. The host operating system is Linux (kernel version 2.4.22). Snort 2.4 is used as the network intrusion detection software [2]. Snort is a commonly used open-source network intrusion detection package that performs signature-based packet inspection. The rule set used consisted of 1751 rules that are organized into 210 chains by Snort. In addition to packet-level signature matching Snort performs various preprocessing tasks on the data. The following preprocessors were enabled during the experiments – frag2, stream4, rpc_decode, bo, telnet_decode. The test set up is shown in Figure 6

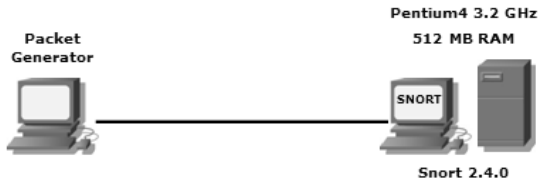

Figure 6.  Test Setup

### B. Traffic Trace

Custom packets were generated using a packet generator with signatures. This was mixed with background traffic and a traffic dump was created. This synthetic traffic profile was used to compare the systems. The *TCPReplay* utility [25] was used to replay packets from another machine to the NIDS. The replayed traffic was speeded up to overload the NIDS program.

### C. Evaluation

The workload and the offered traffic were kept constant during the experiments. Since it is critical that a NIDS should be able to analyze all relevant network packets, the number of attacks detected was monitored. This is required to prove that the signature matching process hasn't dropped any relevant packets.

For evaluation the unnecessary packets sent to the host due to the false positive rate of Bloom Filter, the following analysis was done. Considering that there are $n$ signatures, $k$ hash functions and an $m$ bit vector for the Bloom Filter, the probability of a false positive or the false positive rate can be calculated in a straightforward fashion, given the assumption that hash functions are perfectly random.

After all the elements of S are hashed into the Bloom Filter, the probability that a specific bit is still 0 is

$$\left(1 - \frac{1}{m}\right)^{kn} \approx e^{-kn/m} \qquad (4)$$

Let $p = e^{-kn/m}$. The probability of a false positive is then

$$\left(1 - \left(1 - \frac{1}{m}\right)^{kn}\right)^k \approx \left(1 - e^{-kn/m}\right)^k = (1-p)^k$$

### D. Results

The number of attacks detected matched with a normal NIC and with PEIC, proving that PEIC didn't drop any relevant packets. But the total number of packets analyzed by the host was reduced significantly as shown by Figure 7.

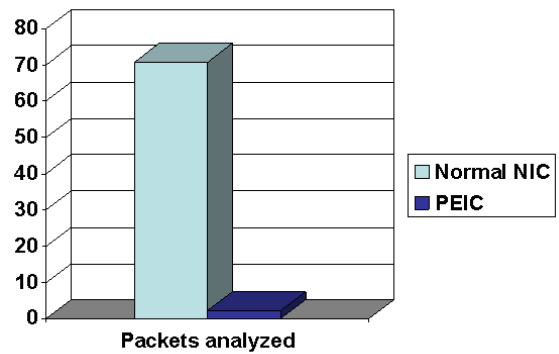

Figure 7.  Percentage of packets analyzed by the host

The plot for the false positive rate for various numbers of hashes and signatures are given below. The bit vector used was of length 2kB. The plot shows that the probability of false positives is very low for up to 2000 signatures.

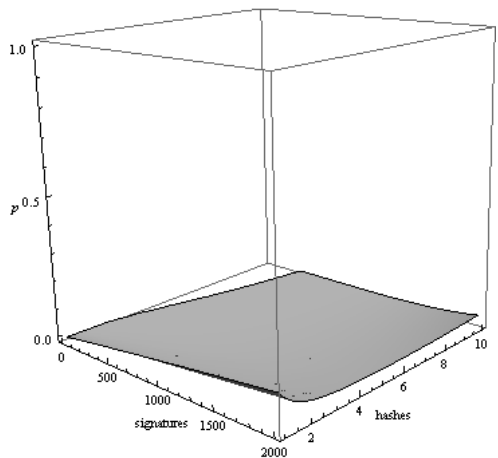

Figure 8. False positive rate for various numbers of signatures and hash functions

## VI. CONCLUSIONS

This paper proposes a network interface card based signature matching and filtering architecture for network intrusion detection systems. The paper argues that hardware based filters incorporated in network interface can significantly increase the system performance and hence increase the quality of detection. It also presents a novel architecture and analyzed its impact on the performance of standard NIDS system. The results show that architecture improves the performance by reducing the excessive interrupts and also by offloading of the costly signature matching operations from the host CPU.